\documentclass[a4paper,11pt]{article}
\usepackage{pos}

\title{Observation of burst activity from SGR1935+2154 associated to first galactic FRB with H.E.S.S.}
\ShortTitle{Observation of SGR1935+2154 with H.E.S.S.}

\author*[a]{D.~Kostunin}
\author[b]{H.~Ashkar}
\author[b]{F.~Sch\"ussler}
\author[c]{G.~Rowell}

\affiliation[a]{DESY, 15738 Zeuthen, Germany}
\affiliation[b]{Institute for Research on the Fundamental Laws of the Universe (IRFU), Commissariat à l'énergie atomique (CEA), Universit\'e Paris-Saclay, F-91191 Gif-sur-Yvette, France}
\affiliation[c]{School of Physical Sciences, University of Adelaide, Adelaide 5005, Australia\\}

\forColl{H.E.S.S.} 

\emailAdd{contact.hess@hess-experiment.eu}

\abstract{
Fast radio bursts (FRB) are enigmatic powerful single radio pulses with durations of several milliseconds and high brightness temperatures suggesting coherent emission mechanism. For the time being a number of extragalactic FRBs have been detected in the high-frequency radio band including repeating ones. The most plausible explanation for these phenomena is magnetar hyperflares. The first observational evidence of this scenario was obtained in April 2020 when an FRB was detected from the direction of the Galactic magnetar and soft gamma repeater SGR1935+2154. The FRB was preceded with a number of soft gamma-ray bursts observed by Swift-BAT satellite, which triggered the follow-up program of the H.E.S.S. imaging atmospheric Cherenkov telescopes (IACTs). H.E.S.S. has observed SGR1935+2154 over a 2 hour window few hours prior to the FRB detection by STARE2 and CHIME. The observations overlapped with other X-ray bursts from the magnetar detected by INTEGRAL and Swift-BAT, thus providing first observations of a magnetar in a flaring state in the very-high energy domain. We present the analysis of these observations, discuss the obtained results and prospects of the H.E.S.S. follow-up program for soft gamma repeaters and anomalous X-ray pulsars.
}

\FullConference{37$^{\rm{th}}$ International Cosmic Ray Conference (ICRC 2021)\\
		July 12th -- 23rd, 2021\\
		Online -- Berlin, Germany}

\begin{document}
\maketitle

\section{Introduction}
Soft Gamma-ray Repeaters (SGR) and Anomalous X-ray Pulsars (AXPs) are associated with highly magnetized neutron stars or magnetars. They generate bursts of emission at irregular time intervals. The crust of the neutron star is thought to break due to the intense shifts of the ultra-strong magnetic field causing the emission of hard X rays and gamma rays. During these short ($\sim$0.1\,s) bursts, the brightness of these objects can increase by a factor of 1000 or more. 
Fast radio bursts (FRBs) are powerful radio pulses of extragalactic origin with a duration of several milliseconds with high brightness temperatures suggesting a coherent emission mechanism~\citep{Petroff:2019tty}. Over the last few years a rapidly growing number of FRBs have been detected in the radio band including repeating ones~\citep{Petroff:2016tcr}. Various theoretical emission and source models were put forward since the first detection of these enigmatic bursts~\citep{2020Natur.587...45Z}. Magnetars are proposed as sources of FRBs~\citep{Popov:2007uv}. For example, some magnetars could produce FRBs through strongly magnetized pulses that interact with the material in the surrounding nebula and produce synchrotron maser emission~(e.g. \citep{Lyubarsky:2014jta}). Some models suggest that repeating FRBs are generated not far from the surface of the magnetar through ultra-relativistic internal shocks and blast waves in the magnetar wind associated with flares~(e.g. \citep{2017ApJ...843L..26B}).

The High Energy Stereoscopic System (H.E.S.S.) is an array of four 12-m and one 28-m Imaging Atmospheric Cherenkov Telescopes (IACTs) located in the Khomas Highland in Namibia at an altitude of 1835~m. It is capable of detecting VHE gamma-rays from energies of a few tens of GeV to 100 TeV. In the past, H.E.S.S. targeted two FRBs~\citep{2017MNRAS.469.4465P, 2017A&A...597A.115H} with several hours to days delay. No significant VHE emission was found from either observations. 

Following its discovery in 2014~\citep{Swift_discovery}, SGR\,1935+2154 has probably become the most burst-active SGR, emitting dozens of X-ray bursts over the past few years~\citep{Lin:2020-I}. SGR\,1935+2154 is associated with the middle-aged galactic SNR\,G57.2+0.8 at a distance of about 6.6~kpc~\citep{Zhou:2020}. In late April and May 2020 SGR\,1935+2154 showed renewed X-ray burst activity culminating with a ``forest of bursts" as detected by BAT onboard the Neil Gehrels {\em Swift} Observatory. Many other X-ray and soft-gamma-ray telescopes ({\em Fermi}-GBM, INTEGRAL, sAGILE, HXMT, Konus-Wind, NICER) also reported sustained bursting activity into late May.

This situation became considerably more interesting with the detection of short, intense radio bursts from the direction of SGR\,1935+2154. Two millisecond-duration radio bursts (FRB\,20200428) were detected in the end of April, the first burst, being a double-peaked one detected by CHIME and STARE2 \citep{SGR1935_CHIME} and the second burst by FAST \citep{FAST_FRB}. The radio energy released under the assumption of an isotropic emission of the bursts at 6.6 kpc is about $10^{34}-10^{35}$\,erg, just below the low end of the extragalactic FRB distribution observed so far. After removing the radio dispersion delay, the timing of the first radio burst appears to line up very well with one of the bright X-ray bursts seen by AGILE~\citep{tavani2020xray}, Konus-Wind~\citep{Ridnaia:2021}, INTEGRAL~\citep{INTEGRAL_BURST_A} and {\em Insight}-HXMT~\citep{HXMT_Bursts,HXMT_Delay}. This coincidence is the first evidence that magnetars are linked to FRBs, or at least a subset of (repeating) bursts. It is also shown that the X-ray bursts overlapping the double-peaked CHIME radio burst have an unusually hard spectrum, and it is suggested that these X-rays and the radio bursts arise from a common scenario \citep{Ridnaia:2021}. Furthermore, the non-thermal nature of the {\em Insight}-HXMT burst~\citep{li2020insighthxmt} points to the production of multi-TeV electrons. Multi-GeV to TeV gamma-ray emission via the inverse-Compton process may then accompany this X-ray emission. 

\section{Observations summary}

\begin{figure}[t]
\centering
\includegraphics[width=1.0\linewidth]{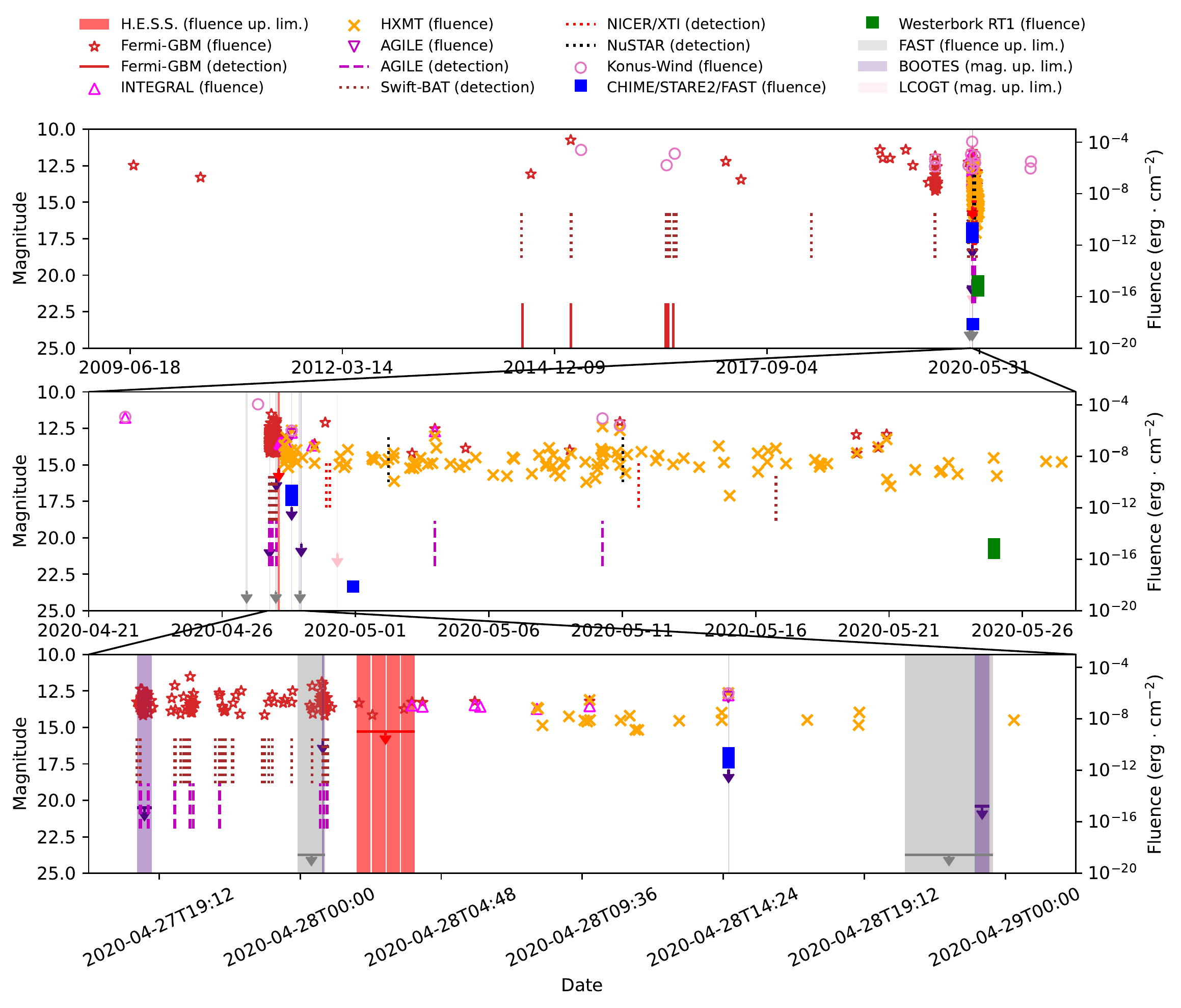}
\caption{SGR\,1935+2154 observations with gamma-ray, X-ray optical and radio telescopes.This plot presents X-ray bursts from the source detected by {\em Fermi}-GBM~\citep{Lin_2020,von_Kienlin_2020}, {\em Swift}-BAT~\citep{Swift-BAT_ATel}, NICER/XTI, NuSTAR~\citep{Borghese_2020}, INTEGRAL~\citep{INTEGRAL_BURST_A}, HXMT~\citep{HXMT_Bursts2,HXMT_Bursts}, AGILE~\citep{AGILE_ATEL,tavani2020xray} and Konus-Wind~\citep{Ridnaia:2021}; radio burst from CHIME~\citep{SGR1935_CHIME}, STARE2~\citep{SGR1935_STAR2}, FAST~\citep{FAST_FRB} and the Westerbork (RT1), Onsala (25m), Toruń (30m) dishes~\citep{Kirsten_2020}. Plot also shows the H.E.S.S., FAST, BOOTES  and LCOGT observations~\citep{2020Natur.587...63L} with upper limits from all shown instruments.}
\label{fig:SGR1935_obs}
\end{figure}

The top  panel of Fig.~\ref{fig:SGR1935_obs} gives an overview of the MWL bursts detected from SGR\,1935+2154 over the years. The middle panel zooms in the 2020 active period. The bottom panel zooms in the period around the H.E.S.S. observations. The observations during the active period are summarized in the following. 

\textbf{Gamma rays and X rays:} AGILE, {\em Fermi}-GBM, INTEGRAL, Konus-Wind, HXMT, NuSTAR, NICER/XTI and {\em Swift}-BAT detected bursts from 2009 until the end of 2020. From the Konus-Wind-detected burst clusters on April 27 2020 we show only the most intense bursting activity.
Four {\em Fermi}-GBM bursts occurred during the H.E.S.S. follow-up observations of the source~\citep{Lin_2020}. INTEGRAL reported the detection of an X-ray burst that coincides with the fourth {\em Fermi}-GBM burst during the last data taking run by H.E.S.S. Therefore the H.E.S.S. observations provide for the first time simultaneous VHE gamma-ray observational data with X-ray bursts emanating from a SGR. An X-ray burst is detected by AGILE, Konus-Wind, HXMT and INTEGRAL simultaneously to the FRB from CHIME and STARE2 (FRB\,20200428). 

\textbf{Radio:} FRB\,20200428 associated with SGR\,1935+2154 is detected by CHIME~\citep{SGR1935_CHIME} and STARE2~\citep{SGR1935_STAR2}. Several X-ray instruments detected coincident X-ray bursts as shown in Fig.~\ref{fig:SGR1935_obs}. Follow-up observations by the FAST radio telescope (Pingtang, China) in the 1.25\,GHz band, did not reveal radio bursts~\citep{2020Natur.587...63L}, however, a weak, highly linearly polarised radio burst was detected on April~30~\citep{FAST_FRB}. Moreover, no radio bursts were detected in the observation campaigns by the  Arecibo, Effelsberg, LOFAR, MeerKAT, MK2 and Northern Cross radio telescopes (not shown in Fig.~\ref{fig:SGR1935_obs})~\citep{Bailes_2021}. Two additional radio bursts separated by $\mathrm{~\sim 1.4 s}$ on May 24, 2020~\citep{Kirsten_2020} were detected following an X-ray burst detected by HXMT during a joint campaign between the Westerbork, Onsala, and Toruń radio dishes. The burst fluences are several orders of magnitude lower than the two bursts detected by CHIME and STARE2. The four FRBs detected from SGR\,1935+2154 thus span around seven orders of magnitude in fluence. No FRBs were detected during the time of the H.E.S.S. data acquisition.

\textbf{VHE gamma-rays:} The H.E.S.S. transients follow-up system triggered Target of Opportunity follow-up observations on SGR\,1935+2154 after the reception of a first {\em Swift}-BAT alert indicating a high-intensity X-ray burst from SGR\,1935+2154 at 2020-04-27 18:26:19.95 UTC. A second {\em Swift}-BAT alert arrived $\sim$6.5 minutes later ~\citep{SWIFT_1_2}. Darkness and visibility constraints only allowed follow-up observations to commence $\sim$7.5 hours later, at 2020-04-28 01:55:00 UTC. The observations lasted 2 hours and consisted of 4 runs taken with {\it wobble} offsets whereby the source is alternately offset by 0.5 deg in opposite directions of Right Ascension and Declination. The positions reported by {\em Swift}-BAT has a 3' uncertainty and is thus fully comprised within the H.E.S.S. field of view of $2.5^\circ$ radius.

\textbf{Optical:} No optical emission has been detected by the BOOTES-3 telescopes observing contemporaneously to the first detected FRB~\citep{2020Natur.587...63L}. No optical emission was seen by LCOGT, other BOOTES telescopes or MeerLICHT optical telescope~\citep{Bailes_2021}.

\begin{figure}[t]
\centering
\includegraphics[height=0.41\linewidth]{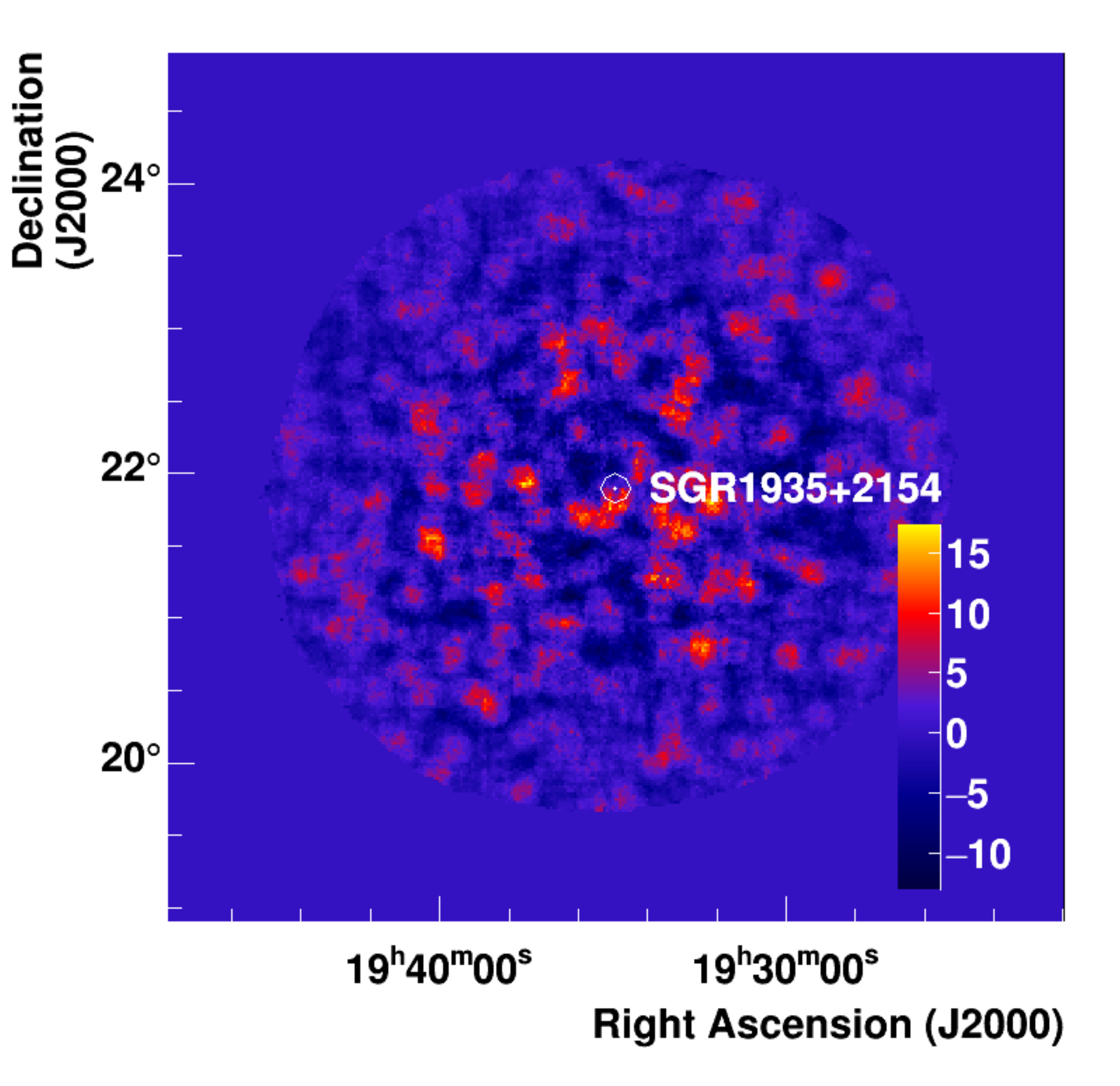}
\includegraphics[height=0.41\linewidth]{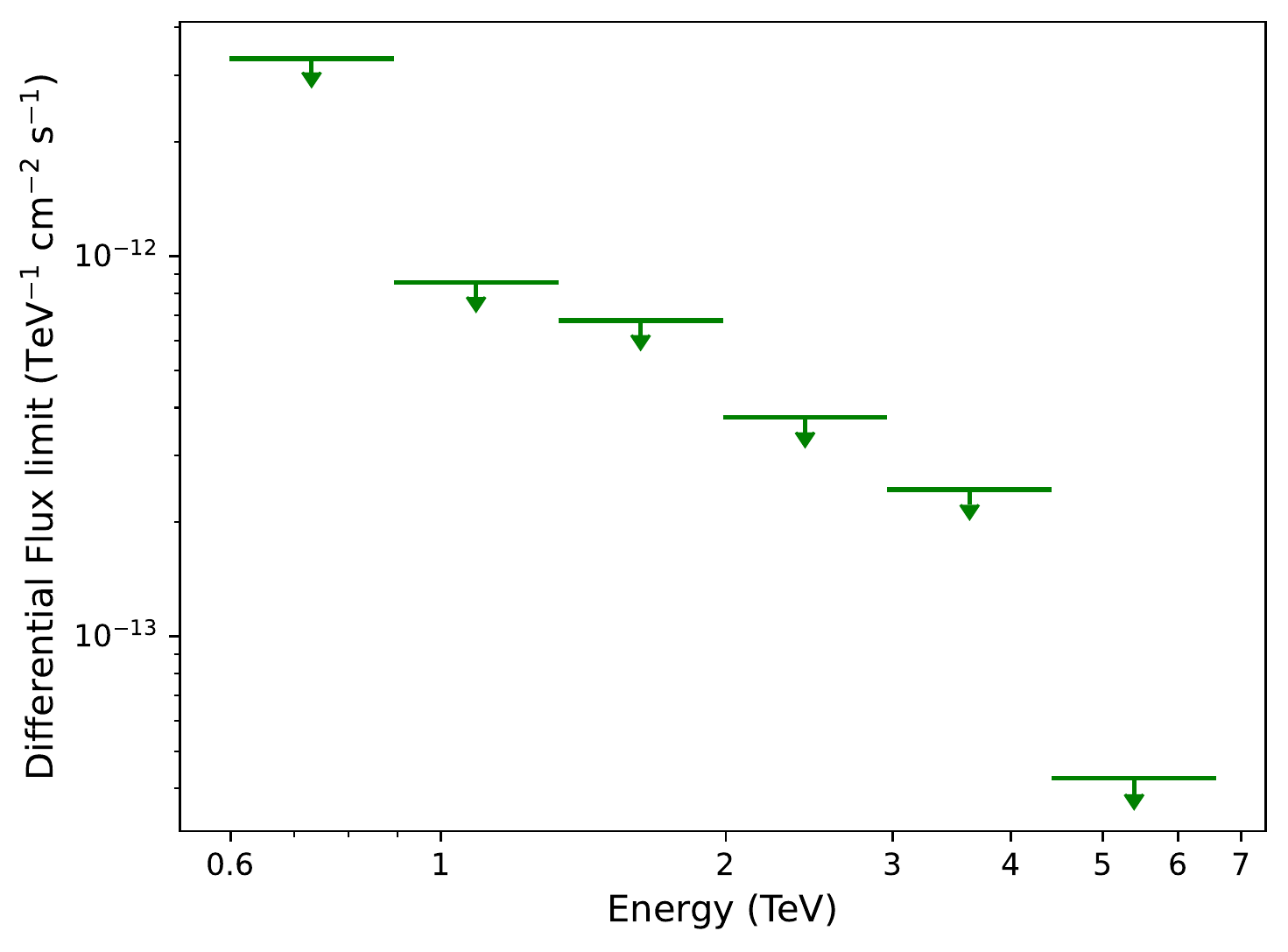}
\caption{\textit{Left:} Excess map computed from the H.E.S.S. observational data taken on SGR\,1935+2154. \textit{Right:} Differential 95\% C.L. upper limits derived from the H.E.S.S. observational data taken on SGR\,1935+2154.}
\label{fig:SGR1935_SIG}
\end{figure}

\section{H.E.S.S. data analysis and results}
\label{sec:SGR1935_hess_analysis}
The data from all four 12\,m telescopes are analysed using the \textit{standard cuts} of the semi-analytical \textit{Model Analysis}~\citep{de-Naurois2009a}. The standard \textit{Ring background} technique~\citep{RingBg} is used to determine the background assuming a radial acceptance and a $\mathrm{0.1\,deg}$ ON region. We obtain a total number of ON events $\mathrm{N_{ON} = 26}$ and OFF events $\mathrm{N_{OFF} = 270}$ in the source region leading to an excess value of~$4.0$ after exposure normalization. We derive the excess map shown in left panel of Fig.~\ref{fig:SGR1935_SIG} with an oversampling radius of $0.1~\mathrm{deg}$. No signal above $5\sigma$ can be found at the position of the SGR or elsewhere in the covered region. We therefore conclude that no significant VHE gamma-ray emission has been detected by H.E.S.S. during the follow-up observations of SGR\,1935+2154.

A low-energy threshold is defined as the energy where the effective area is at least 10\% of its maximum value. Influenced by the relatively high zenith angle of the observations, a value of $\mathrm{E_{thr} = 600 \, GeV}$ is found. Assuming a generic $E^{-2.5}$ energy spectrum we compute 95\% confidence level differential upper limits shown in right panel of Fig.~\ref{fig:SGR1935_SIG} at the position of SGR\,1935+2154 using a Poisson likelihood~\citep{Rolke}. 
Integrating above 600 GeV gives a value of ${\Phi_{\gamma}\mathrm{(E > 600\,GeV)} < 1.5\cdot10^{-12}\,\mathrm{cm}^{-2}\,\mathrm {s}^{-1}}$. The analysis presented in this section has been cross-checked and validated with an independent event calibration and reconstruction analysis~\citep{Parsons2014a}. 

In addition to the standard analysis, we perform the Cumulative Sum, ON-OFF and Exp- tests~\citep{Brun_transient_tools} to search for a variable or transient VHE signal. For that we use the gamma-candidate events that are selected after the cut applied by the \textit{Model Analysis}. No significant variability was detected at minute to hour timescales. Furthermore, a search for gamma candidate doublets arriving within millisecond time windows from the direction of SGR\,1935+2154, which could be associated with magnetar burst is conducted. No such doublets are detected. This search is extended to clusters of gamma candidates around the {\em Fermi}-GBM and INTEGRAL bursts. No gamma-candidate events from the source region are found within less than 9 seconds of any {\em Fermi}-GBM or INTEGRAL burst occurring during our observations. The H.E.S.S. sensitivity for this kind of fast transient phenomena, i.e. assuming detection of gamma-candidate multiplets at millisecond-scale time windows, is at the order of ${\sim 10^{-9}}$\,erg$\,$cm$^{-2}$ depending on the zenith angle of the observations.

\section{Discussion and conclusions}
\label{sec:SGR1935_discussion}

The H.E.S.S. observations described here present the first VHE gamma-ray observations of a magnetar during a high activity phase. Our observations are coincident with four X-ray bursts detected by two instruments ({\em Fermi}-GBM and INTEGRAL) and put stringent upper limits on the VHE emission during the active phase of the magnetar.

The lack of VHE gamma rays is consistent with expectations from regular magnetar bursts~\citep{Kaspi:2017fwg} when thermal emission mechanism is involved. The integral upper limits derived from the two hours of H.E.S.S. observation, assuming a spectral index of $E^{-2.5}$, can be translated into upper limits on the flux $F\mathrm{(E > 600\,GeV)} < 2.4 \times 10^{-12} \,\mathrm{erg} \, \mathrm{cm}^{-2}\,\mathrm{s}^{-1}$. Assuming a distance of 6.6 kpc and isotropic emission we derive a luminosity upper limit  $L\mathrm{(E > 600\,GeV)} < 1.3 \times 10^{34} \,\mathrm{erg}\,\mathrm{s}^{-1}$. This places constraints on persistent VHE emission from SGR\,1935+2154 during the H.E.S.S. observations. The sensitivity to gamma-ray multiplets can be transformed into sensitivity to the isotropic energy from a VHE burst $\mathrm{E_{VHE,iso}\mathrm{(E > 600\,GeV)} \gtrsim 5.2 \times 10^{36} \,erg}$. This sensitivity is higher than the isotropic energies of the FRB bursts detected from the source ($\mathrm{\sim10^{34} \,erg - \sim 10^{35}\, erg}$). However, it can be compared to the energy released in the X-ray domain during the coincident {\em Fermi}-GBM and INTEGRAL bursts ($\mathrm{\sim10^{38} \,erg - \sim 10^{39}\, erg}$), indicating that if there were an isotropic VHE emission related to these X-ray bursts  during the time of H.E.S.S. observations it would have been detected.

The non-detection by H.E.S.S. may suggest that the Inverse Compton process is suppressed in the magnetar surroundings, making the VHE emission too weak to be detected. An explanation for that is that the gamma-ray emission is happening too close to the magnetar surface and pair production and photon splitting result in significant energy losses for the VHE gamma rays leading to strong cutoffs in the MeV to GeV energy range. This flux suppression could be avoided in scenarios where the gamma rays are generated well away from the magnetar's intense magnetic field~\citep{Hu_2019} or in scenarios involving axions~\citep{Archer:2020znv}. In case of detection, the H.E.S.S. observations could therefore also probe the particle transport aspects (such as outflows) in the vicinity of SGR\,1935+2154 during the recent flaring episode. The predicted energy released during a VHE pulse from magnetar nebulae~\citep{Lyubarsky:2014jta} is detectable at distances of hundreds Mpc. 
The derived sensitivity of H.E.S.S. to TeV photons suggests that such VHE bursts would be detectable by H.E.S.S. This motivates further observations of magnetars and FRBs in the VHE domain. 

\section*{Acknowledgements}
\begin{small}
We thank Sergei Popov, Konstantin Postnov and Maxim Pshirkov for fruitful discussions on the interpretation of the observations. 
The H.E.S.S. acknowledgements can be found in:
\url{https://www.mpi-hd.mpg.de/hfm/HESS/pages/publications/auxiliary/HESS-Acknowledgements-2021.html}
\end{small}

\bibliographystyle{JHEP}
\bibliography{sample631}

\clearpage
\section*{Full Authors List: \Coll\ Collaboration}
\scriptsize
\noindent
H.~Abdalla$^{1}$, 
F.~Aharonian$^{2,3,4}$, 
F.~Ait~Benkhali$^{3}$, 
E.O.~Ang\"uner$^{5}$, 
C.~Arcaro$^{6}$, 
C.~Armand$^{7}$, 
T.~Armstrong$^{8}$, 
H.~Ashkar$^{9}$, 
M.~Backes$^{1,6}$, 
V.~Baghmanyan$^{10}$
% ...

\medskip

\noindent
$^{1}$University of Namibia, Department of Physics, Private Bag 13301, Windhoek 10005, Namibia\\
$^{2}$Dublin Institute for Advanced Studies, 31 Fitzwilliam Place, Dublin 2, Ireland\\
$^{3}$Max-Planck-Institut f\"ur Kernphysik, P.O. Box 103980, D 69029 Heidelberg, Germany\\
$^{4}$High Energy Astrophysics Laboratory, RAU,  123 Hovsep Emin St  Yerevan 0051, Armenia\\
$^{5}$Aix Marseille Universit\'e, CNRS/IN2P3, CPPM, Marseille, France\\
$^{6}$Centre for Space Research, North-West University, Potchefstroom 2520, South Africa\\
$^{7}$Laboratoire d'Annecy de Physique des Particules, Univ. Grenoble Alpes, Univ. Savoie Mont Blanc, CNRS, LAPP, 74000 Annecy, France\\
$^{8}$University of Oxford, Department of Physics, Denys Wilkinson Building, Keble Road, Oxford OX1 3RH, UK\\
$^{9}$IRFU, CEA, Universit\'e Paris-Saclay, F-91191 Gif-sur-Yvette, France\\
$^{10}$Instytut Fizyki J\c{a}drowej PAN, ul. Radzikowskiego 152, 31-342 Krak{\'o}w, Poland\\
% ...

%
%
%

\end{document}